\documentclass{aa}  
\usepackage{graphicx}
\usepackage{longtable}
\usepackage{txfonts}
\begin{document}
\title{Toward detection of terrestrial planets in the habitable zone of our closest neighbor: Proxima Centauri
\thanks{Based on observations collected at the European Southern Observatory, Paranal, Chile, programmes 65.L-0428,
66.C-0446, 267.C-5700, 68.C-0415, 69.C-0722, 70.C-0044, 71.C-0498, 072.C-0495, 173.C-0606 and 078.C-0829}}


   \author{M. Endl
          \inst{1}
          \and
          M. K\"urster\inst{2}
          }

   \offprints{M. Endl}

   \institute{McDonald Observatory, University of Texas at Austin,
     Austin, TX 78712, USA\\
              \email{mike@astro.as.utexas.edu}
         \and
             Max-Planck-Institut f\"ur Astronomie, K\"onigstuhl 17, D-69117 Heidelberg\\
             \email{kuerster@mpia-hd.mpg.de}
             }

   \date{Received ; accepted }

 
  \abstract
   {The precision of radial velocity (RV) measurements to detect indirectly planetary companions 
of nearby stars has improved to enable the discovery 
of extrasolar planets in the Neptune and Super-Earth mass range. Detections of extremely low mass planets, 
even as small as 1 Earth mass or below, in short-period orbits now appears conceivable in
ongoing RV planet searches. Discoveries of these Earth-like planets by means of ground-based RV programs will help
to determine the parameter $\eta_{\oplus}$, the frequency of potentially habitable planets around other stars.}
   {In search of low-mass planetary companions we monitored Proxima Centauri (M5V) as part of our M dwarf program. 
In the absence of a significant detection, we use these data to demonstrate the general capability of the RV method 
in finding terrestrial
planets. For late M dwarfs the classic liquid surface water habitable zone (HZ) is located close
to the star, in which circumstances the RV method is most effective. We want to demonstrate that late 
M dwarfs are ideal targets for the search of terrestrial planets with the RV technique.}
   {Using the iodine cell technique we obtained differential RV measurements of Proxima Cen 
over a time span of 7 years with the UVES spectrograph at the ESO VLT. We determine
upper limits to the masses of companions in circular orbits by means of numerical simulations.}
   {The RV data of Proxima Cen have a total rms scatter of $3.1~{\rm m\,s}^{-1}$ and a
period search does not reveal any significant signals. In contrast to our earlier results
for Barnard's star, the RV results for the active M dwarf Proxima Cen are only weakly 
correlated with H$_{\alpha}$ line index measurements. 
As a result of our companion limit calculations, we find that we successfully recover all 
test signals with RV amplitudes corresponding 
to planets with $m \sin i \ge 2 - 3~M_\oplus$ residing inside the
HZ of Proxima Cen with a statistical significance of $>99\%$. Over the same period range, 
we can recover 50\% of the test planets with  masses of 
$m \sin i \ge 1.5 - 2.5~M_\oplus$. Based on our simulations, we exclude the presence of 
any planet in a circular orbit with
$m \sin i \ge 1~M_{\rm Neptune}$ at separations of $a \le 1$~AU.}
   {}

   \authorrunning{Endl \& K\"urster}
   \titlerunning{Terrestrial planets around Proxima Cen}
   \maketitle
%

\section{Introduction}

Over the next decades we will be able to
derive a first estimate of the frequency of stars with a potentially habitable Earth-like
planet. This frequency is usually denoted by the parameter $\eta_{\oplus}$. CoRoT and Kepler are two
space mission that have the capability to detect ``Super-Earths'' and even Earth analogs
in short-period orbits (CoRoT) and at 1~AU (Kepler) using the transit method. 
The astrometry mission SIM Planetquest will achieve the sensitivity to detect Earth-like 
planets around a sample of nearby stars. 
Also, ground-based
Doppler measurements have attained precision levels that make discoveries of
planets with a few Earth masses possible (e.g. Lovis et al.~2006).
Already the very first extrasolar planets found around the pulsar PSR 1257+12 (Wolszczan \& Frail~1992), have
such small masses, that they qualify as terrestrial planets. However, their
formation is likely to have followed a different path than what is currently
envisioned for the formation of terrestrial planets around main sequence stars. 
Besides its obvious astrobiological
implications, and its crucial role for the design and overall costs of any future
TPF/Darwin-type mission, the value of $\eta_{\oplus}$ will also impact our
understanding of the formation of terrestrial planets in general (similar to the
effect the extrasolar giant planets have on giant planet formation models.)

M dwarfs comprise the majority of stars in the solar neighborhood (e.g. Reid et al.~2004) and
once their intrinsic faintness is overcome, they represent attractive targets for 
high precision Doppler surveys. Due to their
lower masses the reflex motion of a planet of a given mass is higher than for a solar mass star.
Thus, it is no surprise that so far the lowest mass extrasolar planets detected by the Doppler method are
all orbiting M dwarfs (Rivera et al.~2005; Udry et al.~2007). The micro-lensing event reported by
Beaulieu et al.~(2006) is also attributed to lensing by a very low-mass planetary companion
to, most likely, an M dwarf host.

In this paper we present 7 years of high precision RV data for our closest neighbor in space: 
the M5V star Proxima Centauri. We demonstrate that with the data in hand we could have 
already detected planets with minimum masses as small as $1 - 2~{\rm M}_{\oplus}$. Constraints for
giant planetary companions to Proxima Cen using HST Fine Guidance Sensor astrometry were reported by
Benedict et al.~(1999) and the companion limits based on less precise RV data from the ESO Coud\'e Echelle 
Spectrometer planet search were presented by K\"urster et al.~(1999). The combination of both 
studies already excluded all companions with (minimum) masses higher than $0.8~{\rm M}_{\rm Jupiter}$ for
the period range $1$ to $600$ days.  


\section{Stellar properties of Proxima Cen}

Proxima Centauri (GJ 551, $\alpha$~Cen~C, HIP~70890) is an M5Ve dwarf and with a distance of d=1.29~pc
the closest star to the Sun. Hipparcos measured a parallax of $772.33\pm2.42$~mas (Perryman et al. 1997).
It is still under debate whether Proxima Cen is actually gravitationally bound to the $\alpha$~Cen~AB binary
(e.g. Wertheimer \& Laughlin 2006). The star exhibits all characteristics of a magnetically active star, 
like coronal X-ray emission (e.g. H\"unsch et al.~1998) and flare activity (e.g. G\"udel et al. 2004).
Benedict et al.~(1998) reported on the photometric variability of Proxima Cen using 4 years of
HST FGS photometry. They find a possible rotational period for Proxima Cen of $\approx84$~days and 
an activity cycle of $\approx1100$~days. Recently Cincunegui et al.~(2007) presented tentative evidence 
for a possible activity cycle with a shorter period of $\approx442$~days.

A crucial parameter for the radial velocity method is the stellar primary mass. Using the V-band
mass-luminosity relationship of Henry et al.~(1999) we derive a mass of $0.108~{\rm M}_{\odot}$ for 
Proxima Cen, while the K-band mass-luminosity relationship of Delfosse et al.~(2000) yields a slightly 
higher mass of $0.12~{\rm M}_{\odot}$. 

\section{Observations and data reduction}

We observed Proxima Cen with the UVES spectrograph at the ESO VLT-UT2 as part of our
ongoing Doppler search for low mass planets around a sample of 40 M dwarfs. We use
a slit width of 0.3 arcsec and image slicer $\#3$ to obtain a resolving power of $R \approx 100,000$.
The 37 orders on the two CCDs cover a wavelength range from $4950$ to
$7040~\AA$.
 
During 76 nights between March 2000 to March 2007 we collected a total of 229 individual spectra
of Proxima Cen typically grouped into three consecutive spectra per night. 
An exposure time of 780 seconds was selected for all spectra. 
The data have an average signal-to-noise ratio of $53.1\pm7.7$ (per pixel).

To allow a simultaneous wavelength calibration, as well as the reconstruction of the
instrumental profile, we employ the standard iodine (I$_{2}$) cell technique for high precision
RV measurements (e.g. Butler et al. 1996). The I$_{2}$ spectrum has
useful reference lines from $\approx 5000$ to $6500~\AA$ that are superimposed  
on the stellar spectrum.

After bias subtraction, flat-fielding, subtraction of inter-order Echelle background,
wavelength calibration, and barycentric correction using the JPL ephemeris DE200 (Standish 1990),
we determine differential RVs with our $Austral$ code as described in Endl et al.~(2000).

\section{RV results}

One result of the data modeling procedure is a formal internal uncertainty of
the RV measurement. This internal uncertainty includes errors due to photon noise, CCD readout noise and 
``algorithmic'' noise in the modeling process to remove all instrumental effects (like spatial and temporal 
variations in the instrumental profile). For the Proxima Cen data we obtained a mean internal error of
$2.34\pm0.28~{\rm m\,s}^{-1}$. 

As discussed in K\"urster et al.~(2003) (from hereon called paper I)
we then formed nightly means of the RV measurements from the typically three (but occasionally more) 
consecutive exposures. Removing this ``clumpiness'' of the data allows for a reliable estimation of the significance
of possible signals using the bootstrap randomization method (see section 6). 
This reduced the total number of data points from 229 to 76. We then subtracted the expected secular
RV acceleration of $0.45~{\rm m\,s}^{-1}\,{\rm yr}^{-1}$, due to the change in perspective to this nearby star, 
from the data (cf. paper I). 
Figure 1 displays our 76 differential RV measurements for Proxima Cen. The velocities
are listed in Table~\ref{rvstab}. 
The data have a total rms of $3.11~{\rm m\,s}^{-1}$, slightly larger than the measurement uncertainties.
The excess RV scatter is thus $2.05~{\rm m\,s}^{-1}$. 

We noted that over the past 2 years the RV scatter of Proxima is greatly reduced. While the data taken before
2005 have an rms scatter of $3.04~{\rm m\,s}^{-1}$, the 16 data points we obtained since then have
a scatter of only $1.36~{\rm m\,s}^{-1}$. Since neither the instrument nor our data reduction algorithm 
was changed over this time, this change is likely instrinsic to the star.

%
   \begin{figure}
   \centering
   \includegraphics[angle=-90, width=9cm]{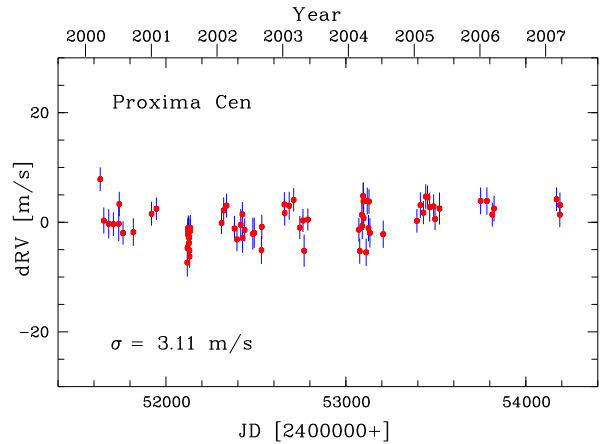}
      \caption{7 years of RV measurements for Proxima Cen using UVES+I$_2$ cell at the ESO VLT. (The secular
acceleration has been subtracted.) The data have a total rms of $3.11~{\rm m\,s}^{-1}$ and an average
uncertainty of $2.34~{\rm m\,s}^{-1}$.
              }
         \label{rvdata}
   \end{figure}
%

\subsection{Period search}

We use the classic Lomb-Scargle periodogram (Lomb~1976 ; Scargle~1982) 
to search for possible periodicities in the 
Proxima Cen RV data. The resulting power spectrum is shown in Fig. 2. 
The highest peak is located at 363.6 days. There is also an intriguing peak 
at longer periods from 2000 to 3000 days,
but such a long period is comparable to our observing time span.
We find no power at the suspected rotation period of $\approx 84$~d, and 
only moderate power at the possible activity cycle period of $\approx 1100$~d.
To estimate the significance of these signals we use the bootstrap randomization
method (e.g. K\"urster et al.~1997). After 10 000 bootstrap randomizations of our
data we find that the 364 d peak has a formal false-alarm-probability (FAP) of just $0.05\%$.
However, this peak is likely produced by the interplay of the strong
1-year peak in the window function (lower
panel in Fig. 2) and the dense data cluster in the middle of 2001. This
cluster is systematically lower than the rest of the points. If we 
reduce the weight of these points by averaging them we find a reduction in
power of the 364-d peak. (The 1-yr peak disappears completely if we remove the 
entire cluster from the time series.) We will discuss the cause of the systematic 
blue-shift of these points in the next section. 
No other peak in the periodogram has a FAP less than $1\%$. 

%
   \begin{figure}
   \centering
   \includegraphics[angle=-90, width=8.5cm]{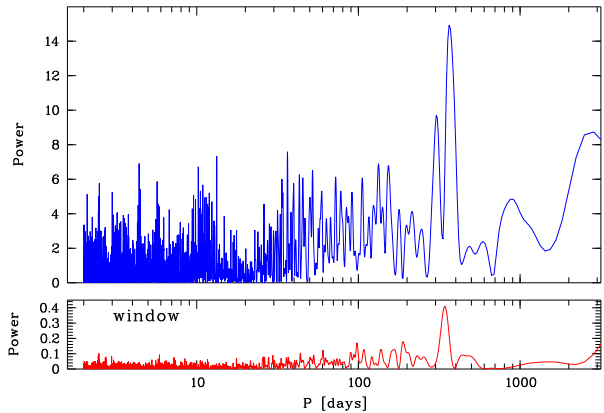}
      \caption{Lomb-Scargle periodogram of the Proxima Cen RV data. The upper panel shows the
power spectrum and the lower panel displays the window function. The 1-year peak in our
window function is quite dominant. Using the bootstrap method to estimate the
significance of the peak shows that no significant signal, except the peak close to 1 year, 
is present in our data.
              }
         \label{lomb}
   \end{figure}
%

\section{H$_{\alpha}$ line index measurements}

As Proxima Cen is a known active flare star we suspected that the excess
scatter might be caused by stellar activity and that we would find a correlation
of the RV data with an activity indicator. In the same fashion as for Barnard's star 
(see paper I for details) we determined a line strength index for H$_{\alpha}$ and, as
a comparison, an index
for a nearby CaI line whose strength should not depend on the activity level of the
star. While for a relatively inactive M dwarf like Barnard's star the chromospheric
emission is seen mostly as a small ``filling in'' of the
H$_{\alpha}$ line core, in the case of Proxima Cen it is seen 
at all times as a strong emission feature above the continuum.
We calculate the line index as the flux inside a region of $\pm 15~{\rm km\,s}^{-1}$ of
the center of the line normalized to the flux of 2 adjacent spectral regions. 
As a check against systematic errors we do the same for the CaI line at 6572.795 $\AA$.
We determine the line index for each individual spectrum and then average the 
values for each night in the same manner as for the RV results. 

Proxima Cen reveals a large amount of variability in H$_{\alpha}$. The average H$_{\alpha}$ line
index is 2.89 with a standard deviation of 0.90 ($31\%$), while the CaI indices is indeed quite constant.
The mean CaI index is $0.57165$ with a standard deviation of $0.0103$ ($1.8\%$).
The amount of variability of the H$_{\alpha}$ index is more than 4 times larger than what we have found in Barnard's star ($7.6\%$).  

However, a period search in the line index data did not find a significant signal, likely due 
to the stochastic nature of the variations (flaring). We find the highest power of 7.13 at a period
of 25.7~days with an estimated FAP of $35\%$.  

   \begin{figure}
   \centering
   \includegraphics[angle=-90, width=9cm]{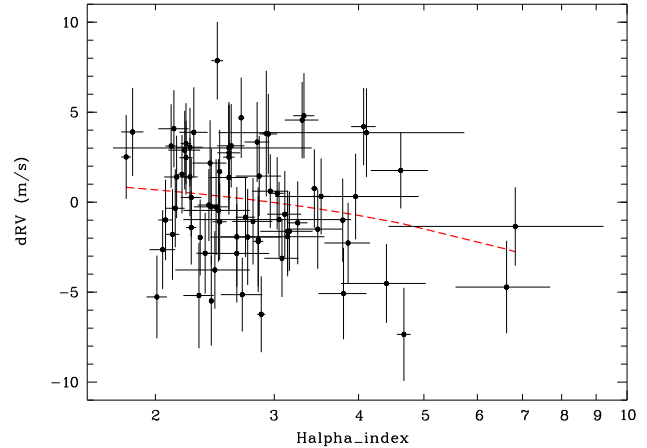}
      \caption{Proxima Cen H$_{\alpha}$ line index measurements as a function of 
RV. The error bar of the line index is the rms of the individual measurements (within
one night) binned to form one data point, in the same manner as the RV data. There appears to be
a weak correlation between the H$_{\alpha}$ index and the RV data. The linear 
correlation coefficient is $-0.23$. A fit to the data is shown as a dashed
line. The residual RV scatter around this slope is $3.03~{\rm m\,s}^{-1}$.}
         \label{halpha}
   \end{figure}
%

Fig.~\ref{halpha} displays the H$_{\alpha}$ line indices as a function of the RV results. We adopt as the error
of the line index the rms of the individual measurements that were averaged.
We find a weak correlation between the H$_{\alpha}$ data and the RV results. 
The linear correlation coefficient is
$-0.23$ resulting in a $4.8\%$ probability that the null-hypothesis, that these two values are not 
correlated, is correct. A linear fit to the data yields a slope of 
$-0.71\pm0.30~{\rm m\,s}^{-1}\,{\rm H}_{\alpha}-{\rm index}^{-1}$. The scatter around this fit is 
$3.03~{\rm m\,s}^{-1}$ (as compared to the overall scatter of 
$3.11~{\rm m\,s}^{-1}$).  At least in a qualitative way this confirms the trend we have found
in paper I: spectra with a higher H$_{\alpha}$ index appear blue shifted. However, 
the correlation for Proxima Cen is much weaker and the slope is much less significant than in the 
case of Barnard's star. (See the detailed discussion in paper I why this might indicate
a general convective redshift for M dwarfs.) It is interesting to note that despite
the fact that Proxima Cen is a much more active star, its activity level does not produce a 
higher RV excess scatter (``jitter'') than in Barnard's star. 

Proxima Cen appears to have been slightly more active during the intensive campaign in summer 2001. 
The average H$_{\alpha}$ line index of this data cluster (16 points) is $3.33\pm0.29$ (error of the mean) while the mean
of the rest of the data is $2.77\pm0.11$ (60 points). This small ($1.4 \sigma$) difference might be the 
explanation why this group of 
measurements is systematically blue-shifted from the rest. 

\section{Limits for planetary companions}

With the absence of a clear RV signal of planets in the Proxima Cen data we pose the question
which type of planets we could have already detected and hence we can exclude.
We compute mass upper limits with a method of injecting and recovering signals into the
original RV data set (Endl et al.~2001). We start with the assumption
that the residual scatter in our data is the best representation of the errors and
use the bootstrap randomization method to determine the statistical
significance of the recovered signal. We intentionally do not correct the RV data with the 
small activity related slope we determined in the previous section, in order to 
determine our detection efficiency also
in the presence of this source of correlated noise. The mass limit for a given period is set at the 
lowest signal amplitude where the signal is recovered at {\it all} trial phase angles
with a significance of $>99\%$. The test signals were generated at 16 different phase angles, 
each shifted by $\pi/8$. We restrict the simulations to circular orbits. Limits derived for
eccentric orbits are typically higher (see e.g. Wittenmyer et al.~2006 for a comparison 
between $e=0$ and $e\le0.6$ limits), but in this case we are particularly interested in planets 
that reside inside the habitable zone (HZ) all the time. An additional constraint is that 
for a successful recovery of the signal the peak in the power spectrum has to be at (or
near) the correct input period. This is especially important for Proxima where the 
1-yr period at the maximum of the window function appears frequently as the dominant signal. 
In a conservative approach we adopt 
the higher value of $0.12~{\rm M}_{\odot}$ for the stellar mass (see section 2).

\begin{figure*}
   \centering
   \includegraphics[angle=-90, width=18cm]{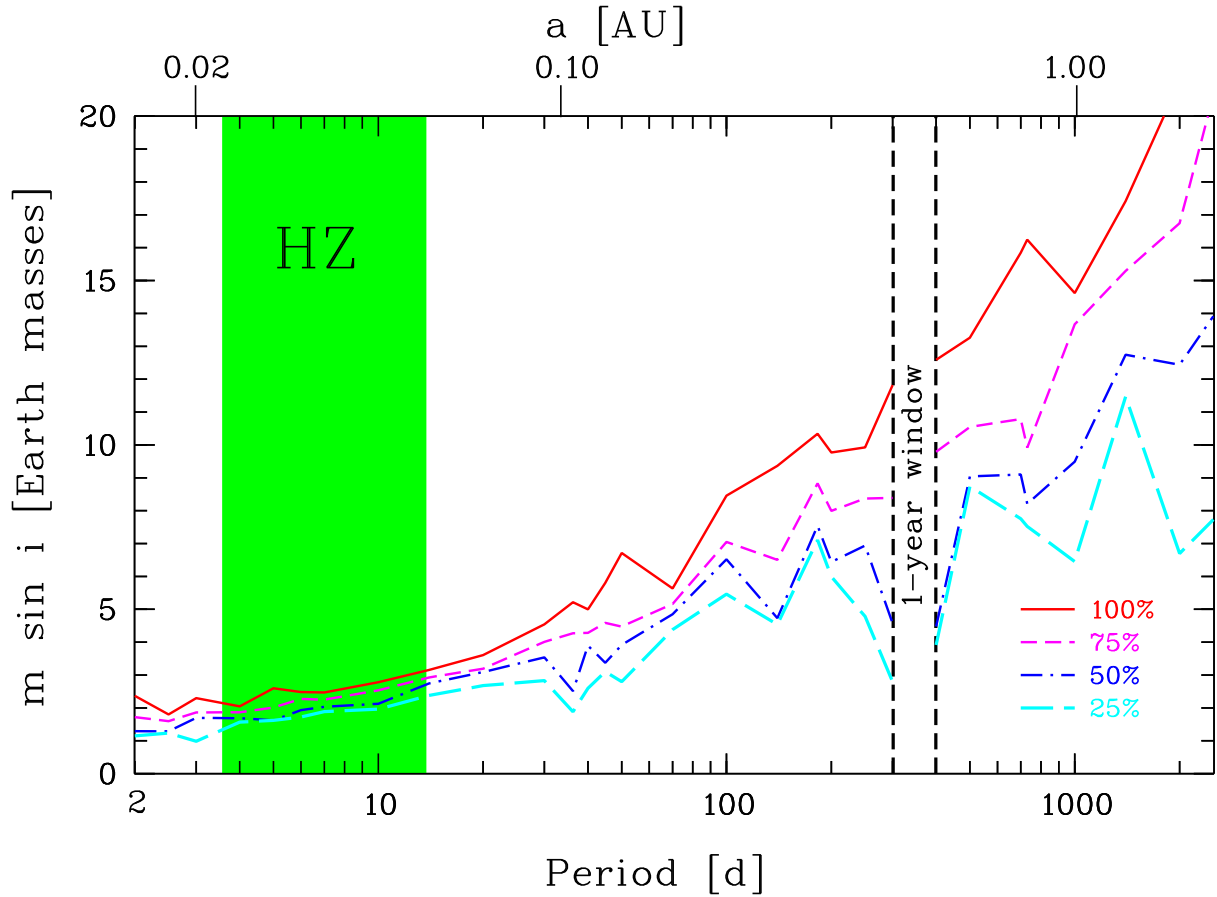}
      \caption{Mass upper limits for planets in circular orbits around Proxima Cen based on our 
numerical simulations. All test signals with amplitudes corresponding to masses on and above the 
solid line were recovered with a $>99\%$ significance. The solid, dashed, dashed-dotted and long-dashed 
lines show the mass range
where we successfully recovered $100, 75, 50$ and $25\%$ of the test signals. The shaded area 
labelled ``HZ'' displays the 
approximate location of the classic liquid water habitable zone (after Kasting et al.~1993).  
The region labelled ``1-year window'' shows the period range that we excluded from our simulations.
              }
         \label{limits}
   \end{figure*}

Fig.\ref{limits} shows the mass upper limits for planets based on this method. The solid line represents
the mass range where all test signals were recovered at (or near) the correct 
period with a significance level of $>99\%$. The shaded area
delimits the classic liquid surface water HZ after Kasting et al.~(1993). For such a late type star 
the HZ is very close to the host star, for Proxima Cen it ranges from 0.022 to about 0.054 AU, corresponding
to orbital periods from 3.6 to 13.8 days. 

We can not determine reliable upper limits in the period range close to 1 year. The fact that
a strong signal at this period is present in the original data leads to the effect that even a 
zero amplitude input signal in phase with the original signal will always lead to a successful
detection. We therefore exclude the region of $300 < {\rm P} > 400$ days from our simulations.
(This excluded part of the mass-period diagram is shown as ``1-year window'' in 
Fig.\ref{limits}.)
  
For periods outside the gap at 1 year we can exclude all planets with
$m \sin i \ge 16~{\rm M}_\oplus$, i.e. planets with minimum masses greater than
Neptune's mass, out to 1 AU. For periods less than 100 days (or $a < 0.21$~AU) we could have
detected all ``Super-Earths'' with $m \sin i \ge 8.5~{\rm M}_\oplus$). For the HZ of Proxima Cen we can rule out
the presence of all planets with $m \sin i \ge 2 - 3~{\rm M}_\oplus$ in circular orbits. In order not to confuse
these upper limits with our actual sensitivity to low mass planets we also show the mass range where 75, 50 and 
25\% of the test signals were recovered with a statistical significance of $>99\%$ (dashed, dash-dotted and long-dashed 
lines in Fig.\ref{limits}). 
This demonstrates that we already had a 50\% chance to detect planets with masses as low as $m \sin i = 1.5~{\rm 
M}_\oplus$ inside the HZ of Proxima Cen and slightly inside the inner edge of the HZ even down to $m \sin 
i \approx 1.2~{\rm M}_\oplus$. 
Over the entire range of periods (with the exception of the 1-yr window) 
we are 100\% (75, 50 and 25\%) complete for RV signals with semi-amplitudes of 
$K = 4.44\pm0.33~{\rm m\,s}^{-1}$ ($3.47\pm0.32, 2.56\pm0.44$ and $2.01\pm0.60~{\rm m\,s}^{-1}$, respectively).

%

\section{Discussion}

We began our VLT/UVES Doppler survey to search for terrestrial planets in the HZ of M dwarfs already in 2000.
We presented first estimates of the survey sensitivity for low mass planets in paper I for Barnard's
star and in Endl et al.~(2003) for Proxima Cen. With the results from this paper we demonstrate that 
the discovery of $m \sin i \approx 1~{\rm M}_\oplus$ is within our grasp. Since sensitivity is a function 
of RV precision,
number of measurements and sampling, adding more points to the existing data string in a pseudo-random fashion,
will allow us to improve the detection sensitivity over time.    

Guedes et al.~(2008) discuss an RV search for terrestrial planets in the HZ of $\alpha$~Cen~B.
They conclude that the detection of a $1.7~{\rm M}_\oplus$ planet will require continuous monitoring of the star 
over at least 3 years and the accumulation of nearly 100 000 independent measurements with a precision of
$\approx 3~{\rm m\,s}^{-1}$. We have shown here that by 
switching targets to Proxima Cen the same feat can be achieved with less than 100 data points. The main
difference being the lower mass of the star as well as the much shorter orbital periods of HZ planets.

When interpreting the mass limits, the reader should bear in mind that we only
considered circular orbits. Limits for planets on eccentric orbits are typically slightly higher (see
e.g. Wittenmyer et al.~2006). Planets with masses above our mass threshold for circular orbits can
still exist around Proxima Cen on eccentric orbits. We also considered only the case of a single
planet. The RV signals of a multi-planet system with several low-mass bodies, is likely to be more
difficult to detect by a pure periodogram analysis (depending on their period spacing and
mass ratios) and requires a significantly larger data sets. Simulations to determine the mass limits for
multiple planets is beyond the scope of this paper.

The question whether terrestrial planets inside the HZ of M dwarfs are suitable for life 
has been discussed extensively in the literature (see e.g. Tarter et al.~2007 
and Scalo et al.~2007 for a review of this subject). At first glance, M dwarf HZ planets did not 
appear to be attractive objects for astrobiology, because
their small semi-major axes would lead to tidal locking into synchronous rotation and presumably 
to atmospheric collapse at the night side of the planet.
However, more detailed climate studies have shown that energy transfer between the day and night side 
of such a planet can prevent atmospheric collapse (Joshi et al.~1997). Furthermore, the long main-sequence
life times of M dwarfs would presumably lead to stable environments on such planets on
very long time scales. Recently, Lissauer~(2007) and Raymond et al.~(2007) discuss the potential of 
habitable M dwarf HZ planets in the context of their formation and in particular water delivery and find
that these planets might be devoid of volatiles and thus less suitable for life.    

One of the drawbacks of the RV technique is the $\sin i$ ambiguity, we measure only minimum mass values while
the true mass remains unknown. But mass is, of course, a crucial parameter for the characterization of a planet.
Low mass planets inside the HZ of M dwarfs have a higher probability to transit their host star, due to 
their smaller semimajor axes. Moreover, the transit depth for a given planet will be deeper for M dwarfs, than
for earlier spectral types (see e.g. Gillon et al.~2007). 
Finding more short periodic transiting Super-Earths around M dwarfs will thus potentially
increase the sample of planets for which we can distinguish their internal composition by putting them
on a mass-radius diagram (e.g. Selsis et al.~2007 ; Valenica et al.~2007).    
 
We have shown that finding terrestrial planets in short periods around M dwarfs is possible even without
an RV precision of $1~{\rm m \, s}^{-1}$ or better. M dwarfs with masses less than $0.2~{\rm M}_\odot$ 
are numerous in the solar neighborhood. However, the majority of them are too faint in the V-band, to be accessed by
traditional Doppler surveys, working in the optical spectral range. A near infrared high resolution spectrograph 
(e.g. Guenther et al.~2006) that can obtain an RV precision of $2 - 3~{\rm m \, s}^{-1}$ would thus be the 
ideal tool to carry out a search
for nearby $m \sin i < 1~{\rm M}_\oplus$ planets.

\section{Conclusions}

   \begin{enumerate}
      \item 
	We present 7 years of high precision RV data for our closest neighbor in space,
the M5V star Proxima Cen, obtained  with UVES + I$_2$ cell at the ESO VLT/UT2. 
We detect no significant periodicities (except close to the 1-yr peak in the
window function) that can be attributed to orbiting companions.  	
      \item 
Using the same set of spectra we measure an H$_{\alpha}$ line index
to estimate the magnetic activity level of Proxima Cen. These line indices show
a large amount of scatter due to flaring activity and are only weakly correlated with the RV results. 
      \item 
Based on numerical simulations we demonstrate that we could have already 
detected all planets with $m \sin i = 2 - 3~{\rm M}_\oplus$ on circular orbits
inside the classic habitable zone of
Proxima Cen (assuming a stellar mass of $0.12~{\rm M}_{\oplus}$). 

   \end{enumerate}

\begin{acknowledgements}
This material is based on work supported by the National Aeronautics and
Space Administration under Grants NNG04G141G, NNG05G107G issued through
the Terrestrial Planet Finder Foundation Science program and Grant
NNX07AL70G issued through the Origins of Solar Systems Program.
We are grateful to the ESO OPC and ESO DDTC for generous allocation of observing time to our 
UVES M dwarf project. A big thank you to the Paranal Science Operations Team who 
performed all the service mode observations. The comments of the anonymous referee  
were very helpful to improve the manuscript. 
We thank Artie P. Hatzes, William D. Cochran, G\"unther Wuchterl and John Scalo 
for interesting discussions
on M dwarf planets. Sebastian Els provided the code for the barycentric correction for the
UVES data and Fr\'ed\'eric Rouesnel and Mathias Zechmeister helped with parts of the spectra extraction.
\end{acknowledgements}

\longtab{1}{
\begin{longtable}{ccc}
\caption{\label{rvstab} Differential radial velocities of Proxima Cen from VLT/UVES}\\
\hline\hline
JD [2 400 000+] & dRV [${\rm m\,s}^{-1}$] & $\sigma$ [${\rm m\,s}^{-1}$] \\
\hline
\endfirsthead
\caption{continued.}\\
\hline\hline
JD [2 400 000+] & dRV [${\rm m\,s}^{-1}$] & $\sigma$ [${\rm m\,s}^{-1}$] \\
\hline
\endhead
\hline
\endfoot
51634.7343 &     7.86 &   2.16 \\
51654.6206 &     0.31 &   2.38 \\
51681.7798 &    -0.28 &   2.66 \\
51707.6289 &    -0.34 &   2.17 \\
51737.5345 &    -0.26 &   3.23 \\
51740.5824 &     3.34 &   2.22 \\
51761.5579 &    -1.96 &   2.11 \\
51818.4977 &    -1.79 &   2.52 \\
51919.8414 &     1.55 &   2.18 \\
51946.8066 &     2.47 &   2.05 \\
52118.5200 &    -7.35 &   2.59 \\
52119.5002 &    -4.72 &   2.57 \\
52120.4995 &    -4.52 &   2.19 \\
52121.5046 &    -1.50 &   2.21 \\
52122.5041 &    -2.27 &   2.23 \\
52123.5158 &    -0.99 &   2.24 \\
52124.5091 &    -1.62 &   2.26 \\
52125.5043 &    -1.92 &   2.20 \\
52126.5086 &    -1.09 &   2.13 \\
52127.5023 &    -3.77 &   2.16 \\
52128.5090 &    -2.84 &   2.24 \\
52129.5172 &    -2.64 &   2.19 \\
52130.5194 &    -5.14 &   2.05 \\
52131.5036 &    -6.23 &   2.11 \\
52134.5201 &    -1.61 &   2.21 \\
52135.5366 &    -1.00 &   2.31 \\
52308.8586 &    -0.16 &   2.00 \\
52320.8412 &     2.17 &   2.38 \\
52336.8520 &     3.05 &   2.19 \\
52380.6817 &    -1.15 &   2.30 \\
52394.7349 &    -3.12 &   2.15 \\
52415.6011 &    -0.46 &   2.85 \\
52424.6250 &     1.45 &   2.23 \\
52425.6540 &    -2.85 &   2.73 \\
52437.5534 &    -1.41 &   2.06 \\
52482.6464 &    -2.09 &   2.91 \\
52490.5975 &    -1.93 &   2.77 \\
52530.5056 &    -5.08 &   2.54 \\
52532.5082 &    -0.84 &   2.21 \\
52657.8318 &     3.25 &   2.25 \\
52659.7972 &     1.70 &   2.32 \\
52684.8259 &     3.01 &   2.56 \\
52709.6935 &     4.08 &   2.14 \\
52743.8400 &    -0.97 &   2.10 \\
52761.5228 &     0.32 &   2.11 \\
52767.9010 &    -5.19 &   2.93 \\
52788.5002 &     0.49 &   2.01 \\
53071.8829 &    -1.35 &   2.18 \\
53076.8916 &    -5.27 &   2.30 \\
53088.8399 &     1.37 &   2.07 \\
53090.8887 &    -0.67 &   2.40 \\
53095.8975 &     4.80 &   2.37 \\
53096.8803 &     3.81 &   3.49 \\
53098.7983 &     0.76 &   2.18 \\
53112.6938 &    -5.49 &   2.48 \\
53118.5595 &     3.86 &   2.46 \\
53125.6977 &    -1.07 &   2.40 \\
53128.6031 &     3.79 &   2.22 \\
53133.5396 &    -1.94 &   2.67 \\
53207.4883 &    -2.18 &   2.50 \\
53394.8511 &     0.26 &   2.17 \\
53413.8344 &     3.13 &   2.32 \\
53430.8431 &     1.76 &   2.11 \\
53443.8731 &     4.69 &   2.23 \\
53453.7691 &     4.56 &   2.12 \\
53465.8306 &     2.74 &   2.64 \\
53487.6901 &     2.89 &   2.18 \\
53495.7335 &     0.61 &   2.05 \\
53520.5247 &     2.50 &   2.92 \\
53748.8611 &     3.90 &   2.45 \\
53783.7137 &     3.88 &   2.50 \\
53813.6861 &     1.40 &   2.18 \\
53823.6115 &     2.51 &   2.33 \\
54170.8614 &     4.20 &   2.15 \\
54188.8877 &     1.40 &   2.29 \\
54189.7161 &     3.12 &   2.32 \\
\end{longtable}
}


\begin{thebibliography}{}

\bibitem[2006]{beaulieu06} Beaulieu, J.-P., Bennet, D.P., Fouqu\'e, P. et al. 2006, Nature, 439, 437

\bibitem[1998]{benedict} Benedict, G. F., McArthur, B., Nelan, E., et al. 1998, AJ, 116, 429

\bibitem[1999]{fritz99} Benedict, G. F., McArthur, B., Chappell, D. W., et al. 1999, AJ, 118, 1086

\bibitem[1996]{butler96} Butler, R. P., Marcy, G. W., Williams, E., McCarthy, C., Dosanjh, P.,
\& Vogt, S.S. 1996, PASP, 108, 500

\bibitem[2007]{cincunegui} Cincunegui, C., Diaz, R. F., \& Mauas, P. J. D. 2007,
A\&A, 461, 1107

\bibitem[2000]{delfosse} Delfosse, X., Forveille, T., Segransan, D., et al. 2000, A\&A, 364, 217

\bibitem[2000]{endl00} Endl, M., K\"urster, M., Els, S. 2000, A\&A, 362, 585 

\bibitem[2001]{endl01} Endl, M., K\"urster, M., Els, S., Hatzes, A.~P., Cochran, W.~D. 
2001, A\&A, 374, 675

\bibitem[2003]{endl03} Endl, M., K\"urster, M., Rouesnel, F., et al. 2003, 
ASP Conference Series, Vol 294, Drake Deming and Sara Seager (eds) 
(San Francisco: ASP) ISBN: 1-58381-141-9, pp. 75-78 

\bibitem[2007]{gillon} Gillon, M., Pont, F., Demory, B.-O., et al. 2007, A\&A, 472, 13

\bibitem[2006]{guenther} Guenther, E.~W., Martin, E.~L., Barrado y Navascues, D., 
\& Laux, U. 2006, Proceedings of ``Tenth Anniversary of 51 Peg-b: Status of and prospects for
hot Jupiter studies, L. Arnold, F. Bouchy \& C. Moutou (eds),
ISBN-2-914601-17.4, p. 326

\bibitem[2004]{guedel} G\"udel, M., Audard, M., Reale, F., Skinner, S. L., \&
Linksy, J.L. 2004, A\&A, 416, 713

\bibitem[2008]{guedes} Guedes, J.~M., Rivera, E.~J., Davis, E., et al. 2008, ApJ, accepted

\bibitem[1999]{henry} Henry, T. J., Franz, O. G., Wasserman, L. H., et al. 1999, ApJ, 512, 864

\bibitem[1998]{huensch} H\"unsch, M., Schmitt, J.~H.~M.~M., Sterzik, M.~F., Voges, W.
1998, A\&ASS, 135, 319 

\bibitem[1997]{joshi} Joshi, M. M., Haberle, R. M., \& Reynolds, R. T.
1997, Icarus, 129, 450

\bibitem[1993]{kasting} Kasting, J. F., Whitmire, D. P., \& Reynolds, R. T. 
1993, Icarus, 101, 108

\bibitem[1997]{mak97} K\"urster, M., Schmitt, J.~H.~M.~M., Cutispoto, G., Dennerl, K. 1997,
A\&A, 320, 831

\bibitem[1999]{mak99} K\"urster, M., Hatzes, A. P., Cochran, W. D., et al. 1999, A\&A, 344, L5

\bibitem[2003]{kuerster} K\"urster, M., Endl, M., Rouesnel, F., et al. 2003, A\&A, 403, 1077

\bibitem[2007]{lissauer} Lissauer, J.~J. 2007, ApJ, 660, 149

\bibitem[1976]{lomb} Lomb, N.~R. 1976, Ap\&SS, 39, 477

\bibitem[2006]{lovis} Lovis, C., Mayor, M., Pepe, F., et al. 2006, Nature, 441, 305

\bibitem[1997]{perryman} Perryman, M.~A.~C., Lindegren, L., Kovalevsky, J., et al. 1997, A\&A, 323, L49

\bibitem[2007]{raymond} Raymond, S.~N., Scalo, J., \& Meadows, V. 2007, ApJ, 669, 606

\bibitem[2004]{reid} Reid, I.~N., Cruz, K.~L., Allen, P., et al. 2004, AJ, 128, 463

\bibitem[2005]{rivera} Rivera, E.~J., Lissauer, J.~J., Butler, R.~P., et al. 2005, ApJ, 634, 625

\bibitem[2007]{scalo} Scalo, J., Kaltenegger, L., Segura, A. G., et al. 2007, AsBio, 7, 85

\bibitem[scargle]{scargl82} Scargle, J.~D. 1982, ApJ, 263, 835

\bibitem[2007]{selsis} Selsis, F., Chazelas, B., Borde, P., et al. 2007, Icarus, 191, 453 

\bibitem[1990]{standish} Standish, E.~M. 1990, A\&A, 233, 252

\bibitem[2007]{tarter} Tarter, J. C., Backus, P. R., Mancinelli, R. L., et al. 2007, AsBio, 7, 30

\bibitem[2007]{udry} Udry, S., Bonfils, X., Delfosse, X., et al. 2007, A\&A, 469, 43  

\bibitem[2007]{valenica} Valencia, D., Sasselov, D.~D., O'Connell, R.~J. 2007, ApJ, 665, 
1413

\bibitem[2006]{wertheimer} Wertheimer, J. G., \& Laughlin, G. 2006, AJ, 132, 1995

\bibitem[2006]{wittenmyer} Wittenmyer, R.~A., Endl, M., Cochran, W.~D., et al. 2006, AJ, 132, 177

\bibitem[1992]{wolszczan} Wolszczan, A., \& Frail, D.~A. 1992, Nature, 355, 145

\end{thebibliography}
\end{document}